\documentclass[twocolumn,preprintnumbers,amsmath,amssymb,superscriptaddress]{revtex4-2}

\usepackage[dvipdfmx]{graphicx}
\usepackage{dcolumn}
\usepackage{bm}
\usepackage{color}
\usepackage{xcolor}
\usepackage{cancel}
\usepackage{hyperref}

\begin{document}

\title{Ising spin ladders of orthopyroxene CoGeO$_3$}

\author{Pavel A. Maksimov}
\affiliation
{Bogolyubov Laboratory of Theoretical Physics, Joint Institute for Nuclear Research, Dubna, Moscow region 141980, Russia}
\affiliation{Institute of Metal Physics, S. Kovalevskoy St. 18, 620990 Ekaterinburg, Russia}

\author{Andrey F. Gubkin}  
\affiliation{Institute of Metal Physics, S. Kovalevskoy St. 18, 620990 Ekaterinburg, Russia}

\author{Alexey V. Ushakov}  
\affiliation{Institute of Metal Physics, S. Kovalevskoy St. 18, 620990 Ekaterinburg, Russia}

\author{Alexander I. Kolesnikov}
\affiliation{Neutron Scattering Division, Oak Ridge National Laboratory, Oak Ridge, TN 37831, USA}

\author{Matthew S. Cook}
\affiliation{Materials Science and Technology Division, Oak Ridge National Laboratory, Oak Ridge, TN 37831, USA}

\author{Michael A. McGuire}
\affiliation{Materials Science and Technology Division, Oak Ridge National Laboratory, Oak Ridge, TN 37831, USA}

\author{G\"unther J. Redhammer}%
\affiliation{Department of Chemistry and Physics of Materials, University of Salzburg, Jakob-Haringer-Strasse 2a, Salzburg A-5020, Austria}

\author{Andrey Podlesnyak}
\affiliation{Neutron Scattering Division, Oak Ridge National Laboratory, Oak Ridge, TN 37831, USA}

\author{Sergey V. Streltsov}  
\affiliation{Institute of Metal Physics, S. Kovalevskoy St. 18, 620990 Ekaterinburg, Russia}

\email[Correspondence and requests for materials should be addressed
to S.S. (email: streltsov.s@gmail.com)]{}

\date{\today}

\begin{abstract}
We present thermodynamic and spectroscopic measurements for an orthopyroxene CoGeO$_3$ with magnetic Co$^{2+}$ ions that form quasi-one-dimensional ladders. We show that non-collinear magnetic order below $T_N$=32 K can be stabilized by a strong local easy-axis anisotropy of $j_\text{eff}=1/2$ moments, which is induced by ligand octahedra distortions. Extraction of a magnetic Hamiltonian from inelastic neutron scattering measurements supports this interpretation and allows us to establish an effective magnetic model. The resulting exchange Hamiltonian justifies CoGeO$_3$ as a realization of an Ising spin ladder compound.
\end{abstract}

\maketitle

\newpage

Many magnetic materials with Co$^{2+}$ ions with $3d^7$ electronic configuration have the ground state characterized by effective total momentum $j_\text{eff}=1/2$~\cite{Abragam}.  In a lattice featuring a common-edge geometry, this can result in strong anisotropy of the exchange interaction, such as Ising spin chains~\cite{coldea2010,Armitage2021,Sid_Co_2020,Coldea_TFIM_2023,Gallegos_2024,Konieczna2025}, exotic states on triangular lattice \cite{Zhong_2019,Gao2022,Nishizawa_2023,Sheng2025,Park_CoI2} and long-sought Kitaev honeycomb model~\cite{KITAEV2006,Liu2018,Sano2018,KimKim21,Winter_Co_2022}.

Cobalt-based pyroxenes with formula ACoX$_2$O$_6$ have recently attracted interest due to their intriguing magnetic properties \cite{Cava_Co_review}. Indeed, in SrCoGe$_2$O$_6$ pyroxene an anisotropic Kitaev interaction was found to be large, on the order of conventional isotropic Heisenberg exchange \cite{maksimov2024}. However, while there are structurally isolated Co chains, the interplay of in-chain anisotropic interactions and frustrated interchain exchange yields a canted antiferromagnetic structure \cite{ding2016,maksimov2024}.

\begin{figure}
    \centering
    \includegraphics[width=0.75\columnwidth]{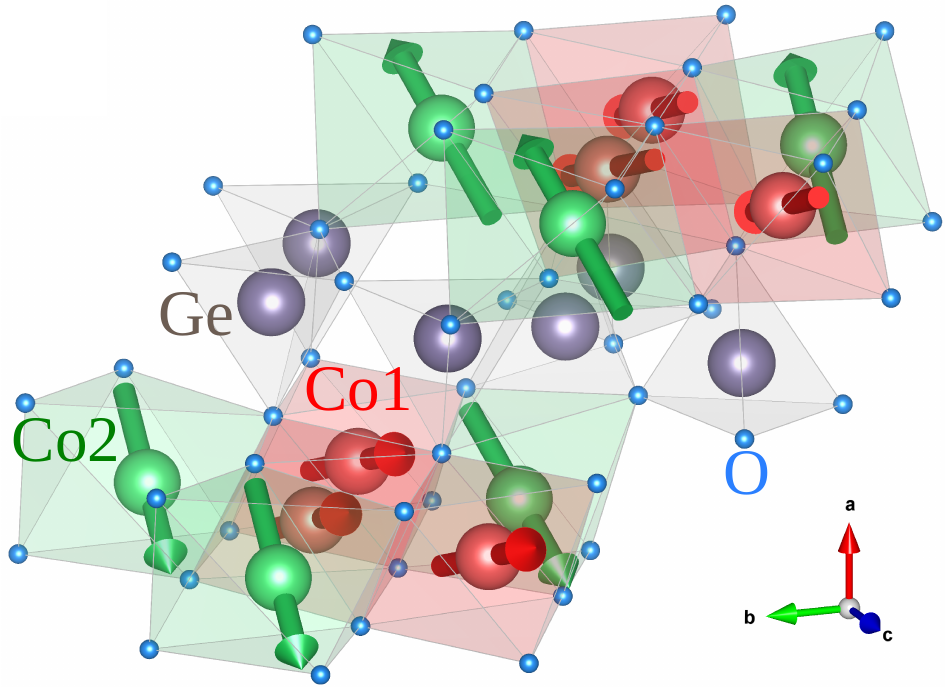}
    \caption{Crystal structure of orthorhombic CoGeO$_3$. Magnetic structure~\cite{redhammer2010a} consists of splayed ferromagnetic ladders, which are running along the $c$ axis and are coupled antiferromagnetically.}
    \label{fig_1}
\end{figure}

\begin{figure}[t]
    \centering
    \includegraphics[width=0.7\columnwidth]{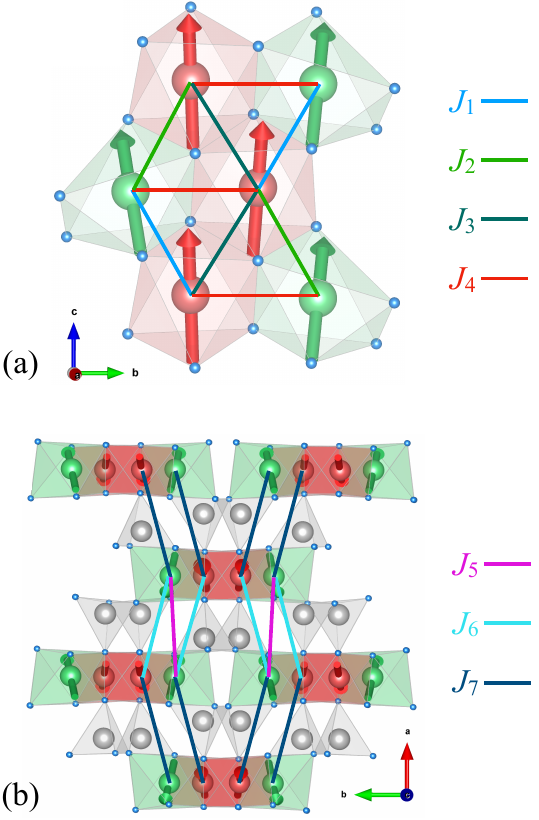}
    \caption{CoGeO$_3$ crystal and magnetic structure in the (a) $bc$ plane (Ge atoms not shown) and (b) $ab$ plane. Relevant exchange paths, ranked by the Co-Co distance, are shown.}
    \label{fig_lattice}
\end{figure}

Moreover, the coupling between chains is expected to be even more important in A-ion deficient pyroxenes, such as CoGeO$_3$~\cite{redhammer2010a,Komarek_2021}, where they are joined together and form spin ladders with triangular motif, see Figs.~\ref{fig_1} and ~\ref{fig_lattice}. This makes them extremely interesting objects for studying low-dimensional magnetism. Indeed, magnetization plateaus were observed not only in the monoclinic polytype of CoGeO$_3$~\cite{guo2021a}, but also recently in CaCoSi$_2$O$_6$~\cite{jin2024}. Furthermore, the spin ladder structure, which is a bridge between one-dimensional and two-dimensional magnetism, exhibits unique ground state and spectral properties \cite{Hida_1991,Scalapino_1992,Barnes_1993,Scalapino_1994,Gopalan_1994,Dagotto_1996,Tsvelik_1996,Giamarchi_book,Vasiliev_2005,Thielemann_2009,Zheludev_2012,Zheludev_2020}. Therefore, a study of cobalt-based spin ladders presents a unique avenue into the interplay of low-dimensional and anisotropic magnetism.

In this Letter, we present detailed experimental and theoretical studies using specific heat, magnetic susceptibility and inelastic neutron scattering (INS) data, as well as \textit{ab initio} and spin-wave calculations for the orthorhombic pyroxene CoGeO$_3$. The results obtained allow us to establish single-ion electronic structure, find parameters for microscopic Hamiltonian, and define the key role of easy-axis anisotropy.

Monoclinic polymorph of CoGeO$_3$ has been studied excessively and was shown to feature a transition to magnetically ordered state at $T_N=35.8$K with non-collinear $C2'/c'$ magnetic structure, strong spin-lattice coupling \cite{redhammer2010a} and distinctive 1/3 magnetization plateau \cite{guo2021a}.
At the same time, orthorhombic CoGeO$_3$ with the $Pbca$ space group \cite{Tauber_1965} is known to exhibit phase transition to a magnetic long-range order at $T_N\approx32$~K \cite{Shaked_1975}. The magnetic structure belongs to the $Pbc^\prime a$  magnetic space group in the parent group setting~\cite{redhammer2010a,Gallego_2016} and corresponds to the splayed ferromagnetic ladders, which are coupled antiferromagnetically. The crystal lattice of orthorhombic CoGeO$_3$, magnetic structure and relevant exchange paths are shown in Figs.~\ref{fig_1} and~\ref{fig_lattice}, where one can observe spin chains running along the $c$-axis. As can be inferred from Fig.~\ref{fig_lattice}(b), the ladders form a square lattice in the $ab$ plane with anticollinear order.

\begin{figure}[t!]
    \centering
    \includegraphics[width=\columnwidth]{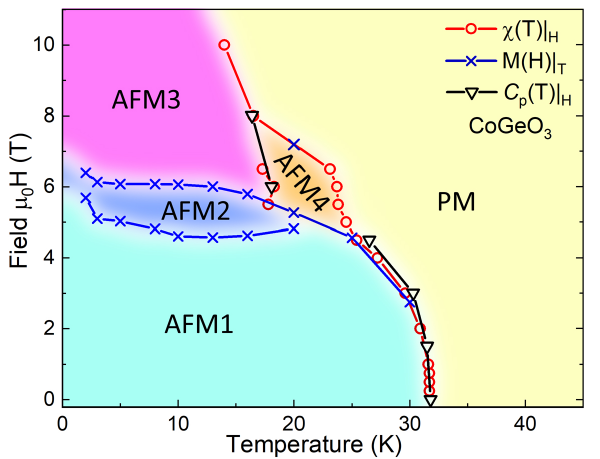}
    \caption{Powder-averaged magnetic phase diagram of CoGeO$_3$ constructed from anomalies in bulk measurements. Symbols mark characteristic anomalies observed in $\chi(T)$, $M(H)$, and $C_p(T)$ for a polycrystalline sample. Shaded regions indicate the approximate stability ranges of four antiferromagnetic phases (AFM1–AFM4). Because the data are from a powder, the phase boundaries represent the onset and completion of transitions across different grain orientations rather than orientation-resolved critical fields. The consistent appearance of the same boundaries in three independent probes demonstrates that these anomalies mark real thermodynamic phase transitions in the polycrystalline sample.}
    \label{fig_pd}
\end{figure}

We carried out specific heat, magnetization and magnetic susceptibility measurements to study the phase transitions of CoGeO$_3$. The measurements were performed using Physical Properties Measurements System (PPMS), Quantum Design, USA, see all the data in the Supplemental Materials~\cite{SM}. Our magnetic susceptibility $\chi(T)\vert_H$, magnetization $M(H)\vert_T$, and heat capacity measurements $C(T)\vert_H$ revealed several magnetic phase transitions induced by an applied magnetic field. In particular, both magnetic susceptibility and heat capacity measured in applied magnetic fields up to $120$~kOe reveal an additional magnetic phase transition at $T_t$ below the N\'{e}el temperature in applied magnetic fields $H=55,~60$, and $65$~kOe \cite{SM}. Moreover, the $M(H)$ curve exhibits two metamagnetic-like transitions at temperatures below $20$~K, three transitions at $T = 20$~K, and a single transition just below the N\'eel temperature $T_N = 32.5$~K \cite{SM}. The resulting phase diagram is shown in Fig.~\ref{fig_pd} where the points correspond to the singularities of the measured macroscopic properties. Note than even though its a polycrystalline sample, the anomalies correspond to physical states for some of the grain orientations. Thus, single-crystal measurements with angle-resolved data may exhibit only some of these features, depending on the field orientation. Nonetheless, the rich magnetic phase diagram revealing four antiferromagnetic phases AFM1, AFM2, AFM3, and AFM4 induced by an external magnetic field and temperature implies the prevalence of anisotropy in magnetic interactions between Co ions in CoGeO$_3$.  Notably, monoclinic polymorph of CoGeO$_3$ also exhibits multiple field-induced phase transitions for some of the field directions due to strong anisotropy \cite{guo2021a}.

\begin{figure}
    \centering
    \includegraphics[width=\columnwidth]{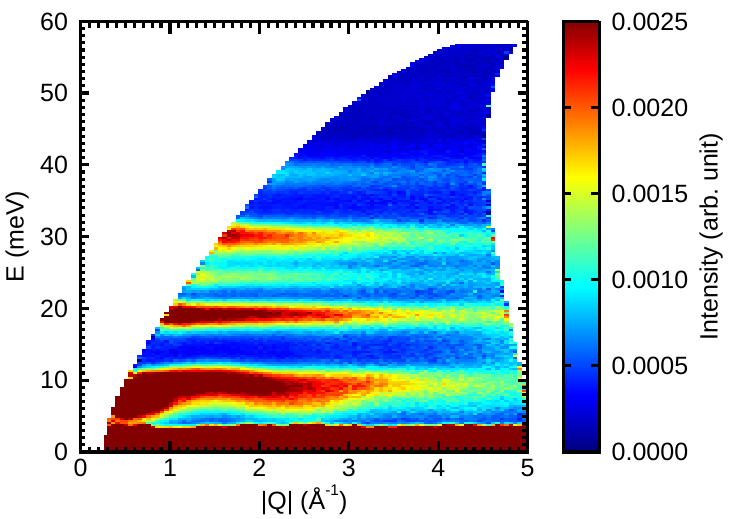}
    \caption{Electronic crystal-field levels from SEQUOIA at $T$=6K, $E_i$=60 meV. One can identify four excitations which correspond to split $j_\text{eff}=3/2$ levels of two inequivalent Co ions.}
    \label{fig_cef}
\end{figure}

In order to confirm the $j_\text{eff}=1/2$ nature of the magnetic ions, we performed high-energy INS using the Fine-Resolution Chopper Spectrometer SEQUOIA~\cite{SEQ} at the Spallation Neutron Source, Oak Ridge National Laboratory. Measurements were carried out at an incident energy of $E_i=60$~meV and a temperature of $T=6$~K. Data analysis was conducted using the Dave software package~\cite{Dave}. We identified four low-lying excitations
above the band of dispersive magnetic response, see Fig.~\ref{fig_cef}. These levels presumably correspond to transitions from the $j=1/2$ to the $j=3/2$ manifold, which is split by trigonal distortions \cite{Winter_Co_2022}, for two inequivalent cobalt sites. The pronounced splitting of the $j=3/2$ level suggests a strong anisotropy of the $g$-factors \cite{Winter_Co_2022}.

We used pseudopotential VASP code~\cite{Kresse1996,Perdew1997} to study crystal-field splitting theoretically. The non-magnetic density functional theory calculations~\cite{Calc_Notes} and Wannier90 package~\cite{Wannier90} were applied to extract single-site non-interacting Hamiltonian. After transformation to the local coordinate system with axes pointing as close to the ligands as possible, this Hamiltonian for the Co $3d$ states was used as a starting point for exact diagonalization calculations for an isolated Co$^{2+}$ ion. In these calculations we take into account both the spin-orbit coupling and correlation effects, which were described in the rotation invariant form~\cite{Georges2013}. 

Taking Hubbard $U=7$ eV, intra-atomic Hund's exchange $J_H=1$ eV~\cite{Maksimov2022Ba,maksimov2024} and the spin-orbit coupling parameter $\lambda=65$ meV~\cite{Abragam} we obtain following energies of the crystal-field excitations: 17 and 20 meV for Co1, and 16 and 93 meV for Co2. One can see that the energies of the first excited states agree rather well with what is observed in the INS data in Fig.~\ref{fig_cef}, while the highest crystal-field excitation for Co2 looks overestimated. The origin of this inconsistency remains unclear — it could arise from a slightly incorrect crystal structure and strong distortions of the Co2O$_6$ octahedra.

In fact, both Co1O$_6$ and Co2O$_6$ octahedra are strongly distorted, the point group is $C_1$ and Co-O bond lengths are 2.09, 2.05, 2.08, 2.11, 2.15, 2.17~\AA~for Co1 and  2.01, 2.03, 2.06, 2.14 2.29, 2.30~\AA~for Co2 at $T=4$ K~\cite{redhammer2010a}, while angles also strongly deviate from 90$^{\circ}$ with $\angle$O-Co-O $\sim$ 81$^{\circ}$ for bonds sharing common O-O edge. 

\begin{figure}
    \centering
    \includegraphics[width=\columnwidth]{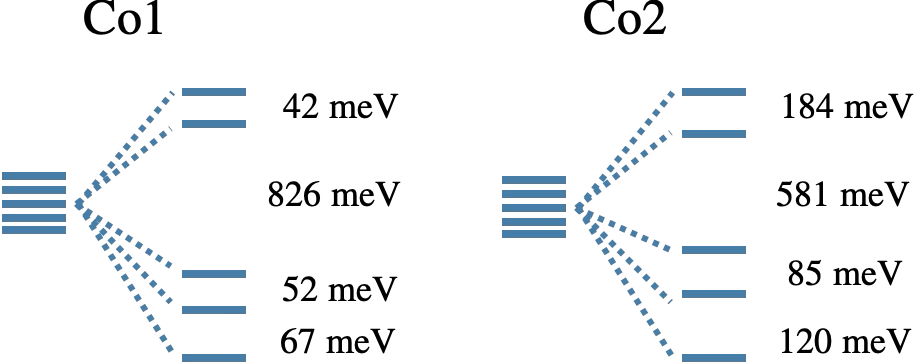}
    \caption{One-electron energies for the two inequivalent Co ions, derived from non-magnetic DFT calculations using Wannier projection. They are distinct from crystal-field excitation energies because they neglect correlation effects.}
    \label{fig_DFT_levels}
\end{figure}

Nonmagnetic DFT calculations show that these strong distortions result in the splitting of all five $3d$ orbitals, as illustrated in Fig.~\ref{fig_DFT_levels}. On the other hand, the resulting energy splittings are comparable to the spin-orbit coupling parameter ($\lambda=65$ meV~\cite{Abragam}). Consequently, the orbital momentum is not quenched. It has long been known that deviations from cubic symmetry not only cause crystal-field splitting but also induce strong anisotropy in the $g$-factor for the effective $j=1/2$ spin Hamiltonian \cite{Lines_1963,Uryu_1976,Abe_1977} This anisotropy can exceed $g_{||}/g_{\perp} \sim 3-4$, see e.g. \cite{Lines_1963,Abragam,griffith1961,Buyers1971,foglio_barberis_2006,Winter2017}, and it is an important factor for analyzing the magnetic excitation spectrum.

\begin{figure*}[t]
    \centering
    \includegraphics[width=2.0\columnwidth]{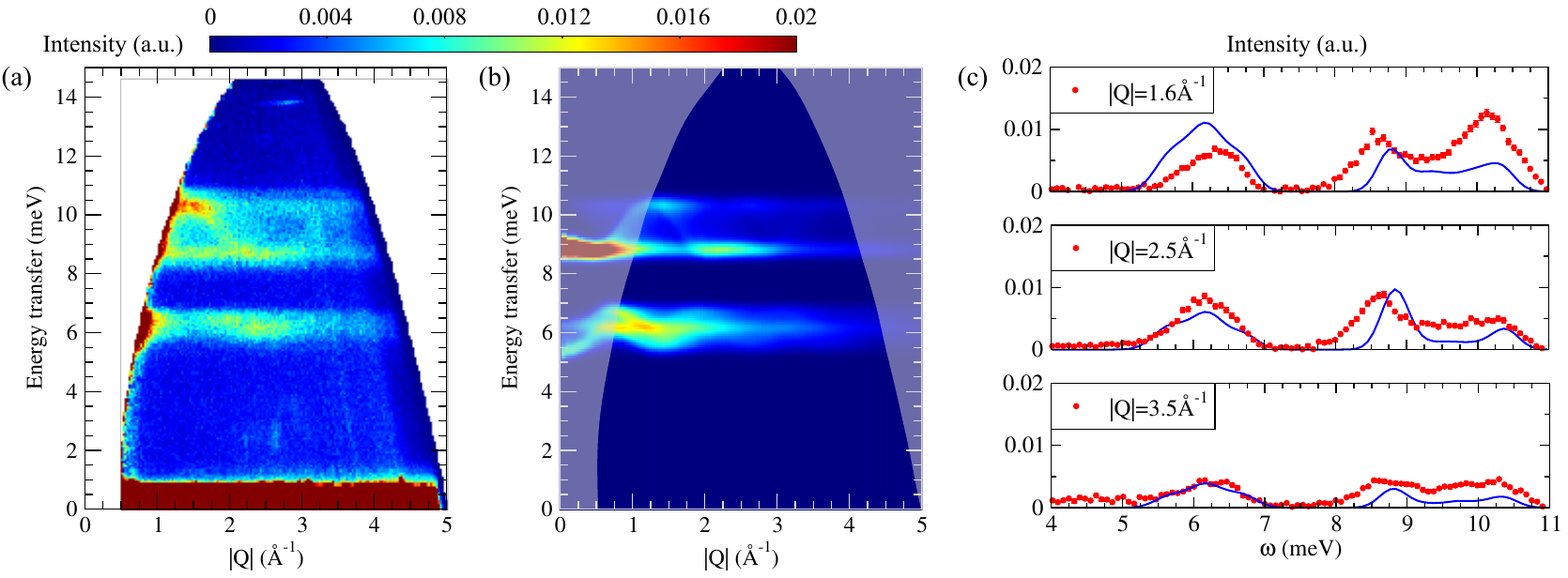}
    \caption{Comparison between the experimental and calculated inelastic neutron spectra in CoGeO$_3$. (a) INS data taken at the CNCS spectrometer at $T=2K$ with incident neutron energy $E_i=15.3$~meV. (b) Linear spin-wave calculation for the powder averaged spectrum for the best-fit set of parameters, see text. (c) Constant-$|Q|$ cuts through the experimental (symbols) and calculated (solid lines) spectral functions for three representative $|Q|$ values.}
    \label{fig_powder}
\end{figure*}

We used data from CNCS spectrometer~ \cite{CNCS1,CNCS2} at Oak Ridge National Laboratory to study the excitations of the pseudospin $j_\text{eff}=1/2$ and extract exchange parameters fitting these data. We performed powder INS with incident energy $E_i$=15.3 meV, at $T=2K$, well below N\'eel temperature. The result is shown in Fig.~\ref{fig_powder}(a) where two bands of spin-wave excitations can be identified.

To perform spin-wave expansion \cite{hp1940}, one needs to ensure that the classical ground state of the Hamiltonian corresponds to the experimental one. The observed magnetic order of CoGeO$_3$ \cite{redhammer2010a} can be a result of the Dzyaloshinskii-Moriya interaction \cite{DM_D}, induced by spin-orbit coupling, which prefers the canted structure~\cite{DM_M}. However, we assume that this structure is the result of distorted octahedra that yields strong $g$-factor anisotropy, in agreement with \textit{ab initio} calculations above. Consequently, local magnetic moments point towards local $g$-factor quantization axes, similar to spin ice \cite{Sondhi_2012} and other anisotropic materials \cite{Zheludev_2025}. Therefore, we suggest that the exchange Hamiltonian can be approximated by the easy-axis $XXZ$ interactions:
\begin{align}
    \mathcal{H}_\text{eff}=\sum_{\langle i,j \rangle_n} J_n \left[\gamma \left( S^{\hat{x}}_i S^{\hat{x}}_j+S^{\hat{y}}_i S^{\hat{y}}_j\right)+ S^{\hat{z}}_i S^{\hat{z}}_j \right],
    \label{eq_ham}
\end{align}
where $\{{\hat{x}},{\hat{y}},{\hat{z}}\}$ refers to the \textit{local} reference frame with $\hat{z}$ pointing along the magnetization direction of the magnetic moment at $\mathbf{r}_i$ and $\gamma$ is the strength of the XY contribution compared to Ising interaction. In order to describe magnetic structure, we need to take into account at least seven exchanges $n=1..7$, see Fig.~\ref{fig_lattice}(b), because one needs to include $J_7$ to stabilize the observed anticollinear interchain structure.

We use SpinW \cite{tothlake} to calculate the dynamical structure factor to simulate the neutron scattering intensity:
\begin{align}
\mathcal{S}(\mathbf{q},\omega)=\mathcal{I}_0 f(q)^2\sum_{\alpha,\beta}\left(\delta_{\alpha \beta}-\frac{q_\alpha q_\beta}{q^2}\right)\mathcal{S}^{\alpha \beta}(\mathbf{q},\omega),
\end{align}
with the dynamical spin correlation function 
\begin{align}
\mathcal{S}^{\alpha \beta}(\mathbf{q},\omega)=\frac{1}{\pi} \text{Im} \int_{-\infty}^\infty dt  e^{i\omega t} \  i\langle \mathcal{T} S^\alpha_\mathbf{q}(t) S^\beta_{-\mathbf{q}}(0)\rangle,
\end{align}
which is calculated with a Gaussian broadening of 0.4~meV to match the width of the experimental peaks, $f(q)$ is cobalt magnetic formfactor. The model \eqref{eq_ham} has eight free parameters that we obtain by minimizing the fit error
\begin{align}
    \chi=\sum_{i,j} \frac{\left(\mathcal{S}^\text{exp}(\mathbf{q}_i,\omega_j) - \mathcal{S}^\text{LSWT}(\mathbf{q}_i,\omega_j) \right)^2}{\delta \mathcal{S}_{i,j}^2},
\end{align}
where the sum is over several momentum transfer values $\mathbf{q}_i$ and energy transfer values $\omega_j>4\text{ meV}$ above the elastic line. Here $\mathcal{S}^\text{exp}(\mathbf{q}_i,\omega_j)$ is intensity from CNCS data, $\delta \mathcal{S}_{i,j}$ is its error, and $\mathcal{S}^\text{LSWT}(\mathbf{q}_i,\omega_j)$ is the linear spin-wave theory (LSWT) calculation. We use MATLAB \texttt{fminsearch} routine \cite{MATLAB} to find the minimum of the fit error (with tolerance $10^{-4}$) and obtain the parameters of the Hamiltonian, shown in Table~\ref{tab:exchanges}, see details in SM~\cite{SM}.

Minimization yields easy-axis anisotropy $\gamma=0.28$, and the scaling factor is given by $\mathcal{I}_0=0.013$. The calculated dynamical structure factor is shown in Fig.~\ref{fig_powder}(b), and the representative constant-$|Q|$ cuts are shown in Fig.~\ref{fig_powder}(c). We see that our model correctly reproduces the position of the magnon peaks but there is a missing intensity at smaller $|Q|$, most likely due to the simplified form of the Hamiltonian. Note that the signs of $J_5$, $J_6$ and $J_7$ in the global reference frames would be flipped as they are responsible for the anticollinear ordering of the neighboring ladders, see Fig.~\ref{fig_lattice}(b).

The model in Table~\ref{tab:exchanges} corresponds to the antiferromagnetically coupled spin ladders with strong easy-axis anisotropy and ferromagnetic intrachain $J_1$ and $J_2$ couplings being the largest in the system.  The diagonal rung coupling $J_3$
is of the same order, while direct rung exchange $J_4$ is subleading.  Significant interladder $J_5$ and $J_7$ interactions yield three-dimensional anticollinear structure. There are intriguing effects, such as phase transitions \cite{Roy_2017} and confinement \cite{Lagnese_2020,Ogino_2021,Ogino_2022}, that can be observed in easy-axis spin ladders, and according to the presented neutron scattering measurements, CoGeO$_3$ falls into this limited class of $S=1/2$ ferromagnetic \cite{Vekua_2003,Kumar_2013} or easy-axis~\cite{Matsuda_2014,Lai2017,Facheris_2024} spin ladders. 

In conclusion, our comprehensive study of orthopyroxene CoGeO$_3$ reveals strong exchange anisotropy among cobalt atoms within its ladder structure, which results in unusual magnetic properties and rich phase diagram.
As magnetization and heat capacity measurements indicate, the antiferromagnetic order of CoGeO$_3$ can be manipulated by moderate magnetic fields. Several intermediate states are observed, which suggests anisotropic exchanges. 

We also observe strong splitting of the $j=3/2$ levels of Co$^{2+}$ ions caused by the distortion of the oxygen octahedra, which is supported by our \textit{ab initio} calculations. These distortions also yield strong anisotropy of exchange interactions \cite{Lines_1963,Winter_Co_2022}, which is in agreement with noncollinear magnetic structure and gapped magnetic spectrum. We confirm this conjecture by extracting parameters of the exchange Hamiltonian from the magnetic dispersion of $j_\text{eff}=1/2$ levels by performing linear spin-wave theory calculations. In fact, we find that the corresponding spin model is essentially Ising-like ferromagnetic ladders that are coupled with antiferromagnetic interactions, which places CoGeO$_3$ in a rather rare class of materials whose magnetic properties are defined by the interplay of anisotropy and low-dimensionality. These findings warrant future work, especially if a single crystal sample is available, to disentangle the nature of the field-induced states and  to place better estimates on the strongly anisotropic exchanges of Co ions.

\begin{table}[t]
 \caption{~Magnetic interactions of CoGeO$_3$ (in meV), as defined in Fig.~\ref{fig_lattice}.} \label{tab:exchanges}
 \begin{ruledtabular}
 \begin{tabular}{ccccccc}
  $J_1$  &  $J_2$ & $J_3$ & $J_4$ & $J_5$ & $J_6$ & $J_7$ \\
         \hline \\[-8pt]
 -4.85&-5.21&-3.48&-0.16&-2.08&-0.09&-0.61 \\
 \end{tabular}
 \end{ruledtabular}
 \end{table}
 
\section*{Acknowledgments}
S.S. thanks Lun Jin and Robert Cava for various stimulating discussions concerning pyroxenes. 
The authors thank Jong Keum for assistance with x-ray measurements.

This research used resources at the Spallation Neutron Source, a DOE Office of Science User Facility operated by Oak Ridge National Laboratory. The beam time was allocated to CNCS and SEQUOIA spectrometers on proposal numbers IPTS-31133 and IPTS-31247, respectively. Heat capacity and magnetization data collection and analysis were supported by the U.S. Department of Energy, Office of Science, Basic Energy Sciences, Materials Sciences and Engineering Division (M.S.C., M.A.M.). X-ray was performed at the Center for Nanophase Materials Sciences (CNMS) at ORNL, which is a DOE
Office of Science User Facility. Calculations of the spin-wave spectrum were supported by Russian Science foundation (project RSF 23-12-00159). The analysis of the phase diagram is supported by the Ministry of Science and Higher Education of the Russian Federation through funding the Institute of Metal Physics.

\section*{Data availability}
INS data supporting the findings of this article are openly available~\cite{Datatof}.


\bibliography{CoGeO3}


\newpage 
\ \
\newpage 
\ \
\newpage
\onecolumngrid
\begin{center}
{\large\bf Supplemental Material:\\
Ising spin ladders of orthopyroxene CoGeO$_3$}\\ 
\vskip 0.35cm
Pavel A. Maksimov,$^{1,2}$~Andrey F. Gubkin,$^2$~Alexey V. Ushakov,$^2$~Alexander I. Kolesnikov,$^3$~Matthew\\
S. Cook,$^4$~Michael A. McGuire,$^4$~G\"unther J. Redhammer,$^5$~Andrey Podlesnyak,$^3$~Sergey V. Streltsov$^2$
\vskip 0.15cm
{\it \small $^1$Bogolyubov Laboratory of Theoretical Physics, Joint Institute for Nuclear Research, Dubna, Moscow region 141980, Russia}\\
{\it \small $^2$Institute of Metal Physics, S. Kovalevskoy St. 18, 620990 Ekaterinburg, Russia}\\
{\it \small $^3$Neutron Scattering Division, Oak Ridge National Laboratory, Oak Ridge, TN 37831, USA}\\
{\it \small $^4$Materials Science and Technology Division, Oak Ridge National Laboratory, Oak Ridge, TN 37831, USA}\\
{\it \small $^5$Department of Chemistry and Physics of Materials, University of Salzburg, Jakob-Haringer-Strasse 2a, Salzburg A-5020, Austria}\\
{\small (Dated: \today)}\\
\end{center}
\vskip 0.2cm \
\twocolumngrid

\setcounter{page}{1}
\thispagestyle{empty}
\makeatletter
\renewcommand{\c@secnumdepth}{0}
\makeatother
\setcounter{section}{0}
\renewcommand{\theequation}{S\arabic{equation}}
\setcounter{equation}{0}
\renewcommand{\thefigure}{S\arabic{figure}}
\setcounter{figure}{0}
\renewcommand{\thetable}{S.\Roman{table}}
\setcounter{table}{0}
{\bf {Supplementary Note 1: Sample synthesis and x-ray diffraction}}
\smallskip

\begin{figure}[htb!]
    \centering
    \includegraphics[width=1.0\columnwidth]{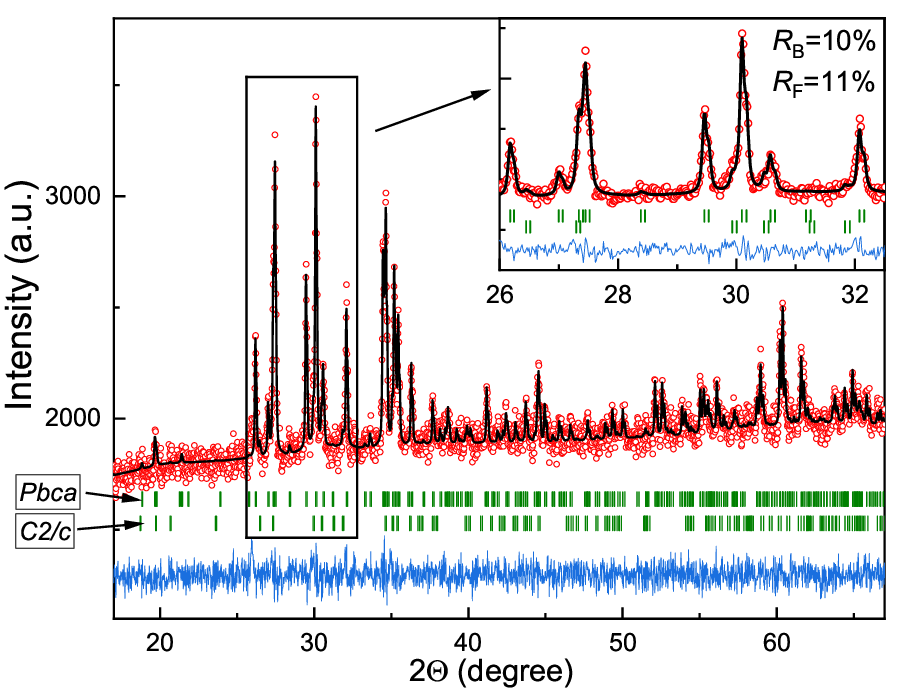}
    \caption{Rietveld refinement of the x-ray diffraction pattern measured at room temperature. Observed, calculated profiles and difference curves are represented by red circles, a solid black line through the symbols, and a blue line at the bottom, respectively. The first line of bars below the patterns represents Bragg peaks positions of the orthorhombic phase of CoGeO$_3$ while the second line corresponds to the monoclinic impurity phase of CoGeO$_3$.}
    \label{x-ray}
\end{figure}

\noindent

The powder sample was synthesized using technique described in Ref.~\cite{redhammer2010a}. The x-ray diffraction measurements were performed at room temperature to carry out structural characterization and to check the phase purity of the CoGeO$_3$ powder sample. The measurements were conducted on a PANalytical X$^\prime$Pert Pro MPD equipped with an X$^\prime$Celerator solid-state detector. The x-ray beam was generated at $45$~kV/$40$~mA, and the wavelength was set at $\lambda=1.5418$~\AA\ (Cu K$\alpha$ radiation). The step size was 0.017$^{\circ}$.  Rietveld refinement has been performed using the orthorhombic crystal structure model (space group $Pbca$) previously reported for CoGeO$_3$~\cite{Tauber}. It was found that this model describes well most of the Bragg peaks in the x-ray diffraction pattern. However, a weak contribution of the monoclinic CoGeO$_3$ impurity phase SG $C2/c$ ($\approx 9$~wt.\%) was observed in the x-ray diffraction data. The best-fit result is shown in Supplementary Figure~{\ref{x-ray}}. The refined unit cell parameters of the main CoGeO$_3$ phase are as follows $a=18.8505(10)${\AA}, $b=9.0274(5)${\AA}, $c=5.3814(3)${\AA}.

\bigskip
\noindent

\noindent
{\bf {Supplementary Note 2: Magnetization measurements}} 
\smallskip

\begin{figure*}[htb!]
    \centering
    \includegraphics[width=1.8\columnwidth]{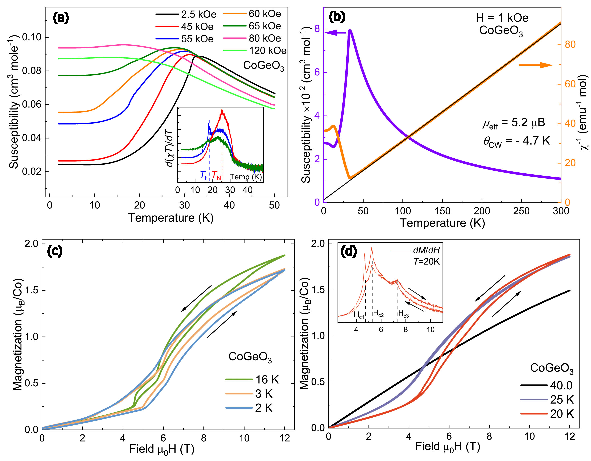}
    \caption{(a) Magnetic susceptibility as a function of temperature for various values of magnetic field. (b) Magnetic susceptibility as a function of temperature for $H=1$ kOe and inverse magnetic susceptibility. (c),(d) Isothermal magnetization as a function of the magnetic field measured at various temperatures. Inset shows susceptibility as a function of magnetic field.}
    \label{magnetic}
\end{figure*}

\noindent

For magnetization measurements, $46.1$~mg of CoGeO$_3$ powder sample were packed and loaded in a Quantum Design Physical Property Measurement System (PPMS) Dynacool instrument. Temperature-dependent magnetization data were collected under field cooling conditions from 50~K to $2$~K and sweep mode at a number of constant fields $2.5$, $5$, $7.5$, $10$, $20$, $30$, $40$, $45$, $50$, $55$, $60$, $65$, $80$, $100$, and $120$~kOe. Additional measurements were performed in a constant field of $1$~kOe in a wide temperature range $1.8-300$~K under field cooling conditions to provide high-temperature susceptibility data for a Curie-Weiss (CW) fit. Magnetization $M(H)$ was measured as a function of the applied magnetic field at constant temperature on the same sample in the field range $0 \rightarrow 120  \rightarrow -120  \rightarrow 0$~kOe at a number of temperatures $2$, $3$, $5$, $8$, $10$, $13$, $16$, $20$, $25$, $30$, and $50$~K.

Supplementary Figure~\ref{magnetic}(a) represents magnetic susceptibility curves measured as a function of temperature under the FC protocol in applied magnetic fields up to $120$~kOe. As one can see,  $\chi(T)$-curves reveal a well-defined anomaly at the N\'{e}el temperature $T_N=32.5$~K that moves towards low temperatures when the external magnetic field increases. When the external magnetic field reaches $55$~kOe, a second anomaly emerges at the temperature $T_t=17.7$~K implying the field-induced intermediate magnetic phase existing in the narrow temperature range between $T_t$ and $T_N$. A future increase of the magnetic field $65$~kOe suppresses the intermediate magnetic phase when the magnetic subsystem evolves towards a field-induced ferromagnetic state. 

The approximation of the inverse magnetic susceptibility $\chi^{-1}(T)$ with the Curie-Weiss law is shown in Supplementary Figure~\ref{magnetic}(b). It can be seen that the linear CW fit describes well the $\chi^{-1}(T)$-curve in a wide temperature range above $50$~K. The estimated effective magnetic moment was found to be as high as $\mu_\text{eff}=5.2~\mu_B$ while the Curie constant $\theta_{CW}=-4.7$~K. The estimated effective magnetic moment is substantially higher than the theoretical spin-only value $\mu_\text{eff}=3.87$~$\mu_B$ that one can expect for the $\mathrm{Co}^{2+}$ ion. The enhanced effective moment was previously reported for the monoclinic phase of CoGeO$_3$ and was ascribed to an orbital contribution that gives rise to an Ising-like antiferromagnetic state at low temperatures~\cite{Guo}.

Magnetization curves measured at low temperatures below $20$~K (see Supplementary Figure~\ref{magnetic}(c)) reveal two successive metamagnetic-like transitions and still remain far from full saturation with the magnetization value $M(2~\mathrm{K}, 120~\mathrm{kOe})=1.7~\mu_B/\mathrm{Co}$. The magnetization curve measured at $T=20$~K (see Supplementary Figure~\ref{magnetic}(d)) exhibits complicated behavior with three successive metamagnetic transitions, as evidenced by the $dM/dH(20~\mathrm{K})$ curve (see Supplementary Figure 2(d)). Further heating results in a single-transition behavior at $T=25$~K and Brillouin-like behavior  above the N\'{e}el temperature at $T=40$~K. The fit of the $M(H)$ curve measured in the paramagnetic state at $T=40$~K with the Brillouin function provides a good fit quality and $J\approx2$ that is again larger than the spin-only value $S=\frac{3}{2}$. While our model from the main text does not reproduce magnetization data in Fig.~\ref{magnetic}(c) adequately, we should note that preliminary calculation shows that saturation field for $H||c$ (around 3 T) is much smaller that critical field for $H\perp c$ (around 25 T), which might provide insight into the nature of multiple observed field-induced states.

Note that the absence of singularity at 36 K \cite{redhammer2010a} and low-temperature field-induced transition around 8 T \cite{guo2021a} implies that monoclinic phase impurity does not contribute to the magnetic properties of our sample.

\bigskip
\noindent

\noindent
{\bf {Supplementary Note 3: Heat capacity}}
\smallskip

\noindent

\begin{figure}[htb!]
    \centering
    \includegraphics[width=0.95\columnwidth]{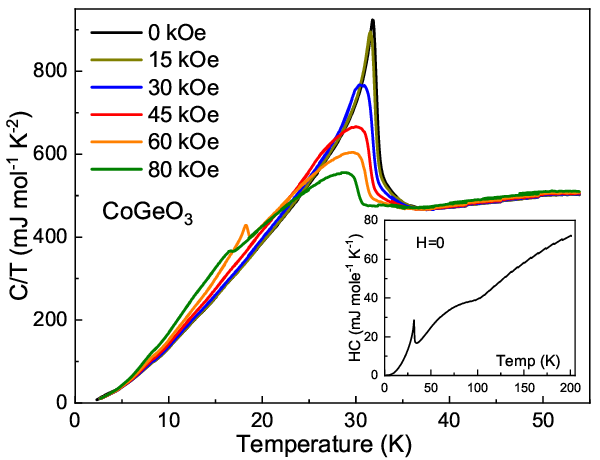}
    \caption{Heat capacity as a function of temperature $C(T)/T$ in applied magnetic fields up to $80$~kOe. Zero-field curve $C(T)$ is shown in the inset}
    \label{HC}
\end{figure}

Heat capacity was measured using a Quantum Design PPMS. For the heat capacity measurement, approximately 2.77~mg of CoGeO$_3$ powder was mixed with $\sim$1~mg of Apiezon N grease. The addenda contribution was determined separately using $\sim$1~mg of Apiezon N grease deposited directly onto the specific heat puck. The heat capacity of the CoGeO$_3$ sample was obtained by subtracting the measured addenda contribution from the total heat capacity.

Supplementary Figure~\ref{HC} represents the heat capacity data measured in applied magnetic fields up to $80$~kOe. It can be seen that the $C(T)$ curve measured in zero magnetic field exhibits $\lambda$-type anomaly at the  N\'{e}el temperature $T_N\approx32$~K accompanied with a broad Schottky-type anomaly above the N\'{e}el temperature. The application of the external magnetic field results in gradual suppression of the $\lambda$-type anomaly and its shift to lower temperatures, as one can expect for the antiferromagnetic phase. In external magnetic fields $60$~kOe, the second anomaly emerges on the $C(T)/T$ curve indicating an order-order type magnetic phase transition below the N\'{e}el temperature. The emergence of the intermediate magnetic phase in the applied magnetic field of  $60$~kOe is consistent with the magnetic susceptibility data reported in Supplementary Note 2.

\bigskip
\noindent

\noindent
{\bf {Supplementary Note 4: First-principles calculations}}
\smallskip

\noindent

\begin{figure}[t!]
    \centering
    \includegraphics[width=0.95\columnwidth]{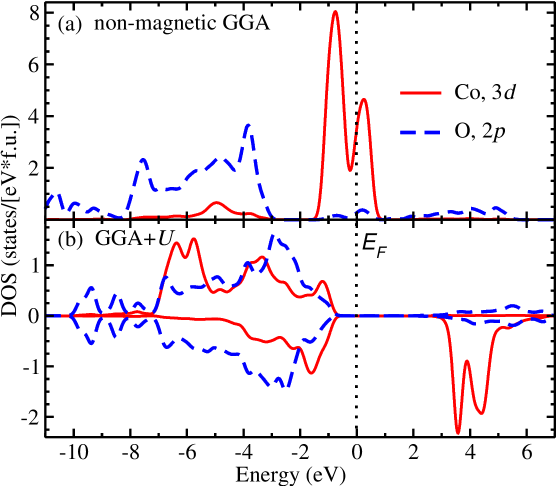}
    \caption{Partial densities of states of CoGeO$_2$ obtained within (a) non-magnetic GGA and (b) magnetic GGA+$U$ approximation. The Fermi level is in zero.}
    \label{dos}
\end{figure}

The partial densities of states (PDOS) of CoGeO$_3$ are presented in Supplementary Figure~\ref{dos}. One can see from non-magnetic GGA calculations, that the system exhibits typical behavior for 3$d$ transition metal oxides: relatively narrow Co-3$d$ bands are located near the Fermi energy, spanning the range from -1.5 to 1 eV, while broad O-2$p$ states lie at lower energies (-11 to -3 eV) and they are well-separated ($\sim$1.5 eV) from the Co-3$d$ states. The Co-3$d$ density of states features a two-peak structure corresponding to the $t_\textrm{2g}$ and $e_\textrm{g}$ levels, suggesting a crystal field splitting of approximately 0.8 eV in CoGeO$_3$.

With the inclusion of on-site Hubbard $U$ and Hund's coupling $J_\textrm{H}$, the system transforms to an insulating state with an energy gap $E_\textrm{g}$ = 3.6~eV (we used experimental magnetic order in these calculations). The Co-3$d$ occupation number is 7.11 electrons and the local magnetic moment of 2.8 $\mu_\textrm{B}$ confirm the Co$^{2+}$ valence state. The barycenter of the occupied 3$d$ levels shifts downward to approximately -4 eV, enhancing hybridization with the O-2$p$ states. Additionally, within the unoccupied $e_\textrm{g}$ states, a splitting of 0.9 eV is observed between the lower-energy $3z^2-r^2$ and higher-energy $x^2-y^2$ orbitals.

\bigskip
\noindent

\noindent
{\bf {Supplementary Note 5: Spin-wave theory calculations}}
\smallskip

\begin{figure}
    \centering
    \includegraphics[width=\columnwidth]{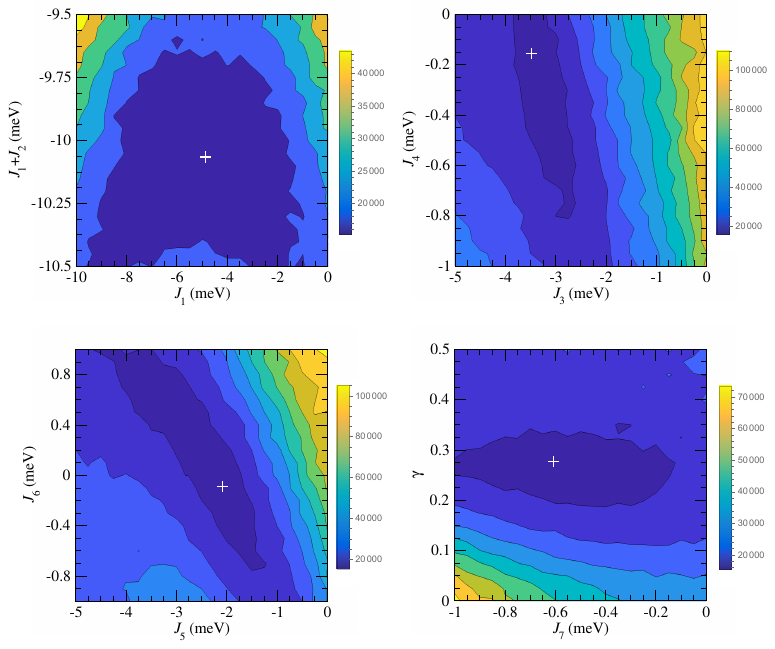}
    \caption{Intensity plots of fitting error $\chi$ as a function of exchange integrals. White cross marks the parameters set from the main text that corresponds to obtained minimum.}
    \label{fig_fit_error}
\end{figure}

Since a transition to a long-range magnetic ordered state is observed in heat capacity and magnetic susceptibility measurements, see Figs.~\ref{magnetic} and \ref{HC}, we employ spin-wave theory to describe the excitations. 

Holstein-Primakoff transform \cite{hp1940}
\begin{align}
S^x&\approx\sqrt{\frac{S}{2}} \left(a  +a^\dagger\right),\nonumber\\
S^y&\approx\frac{1}{i}\sqrt{\frac{S}{2}} \left(a -a^\dagger \right),\nonumber\\
S^z&=S-a^\dagger a,
\label{eq_HP}
\end{align}
yields the bosonic Hamiltonian that can be diagonalized to obtain magnon spectrum and neutron scattering intensity. We use SpinW to perform spin-wave calculations and to obtain dynamical structure factor that can be compared to the CNCS data.

In order to obtain exchanges of the Hamiltonian, we searched for the global minimum of the fitting error of the powder scattering data. The fitting error is shown in Fig.~\ref{fig_fit_error} for several cuts of the multidimensional parameter phase space. One can see that the obtained parameter set (marked with the white cross) does in fact correspond to a minimum of the error. 

\begin{figure}
    \centering
    \includegraphics[width=0.95\columnwidth]{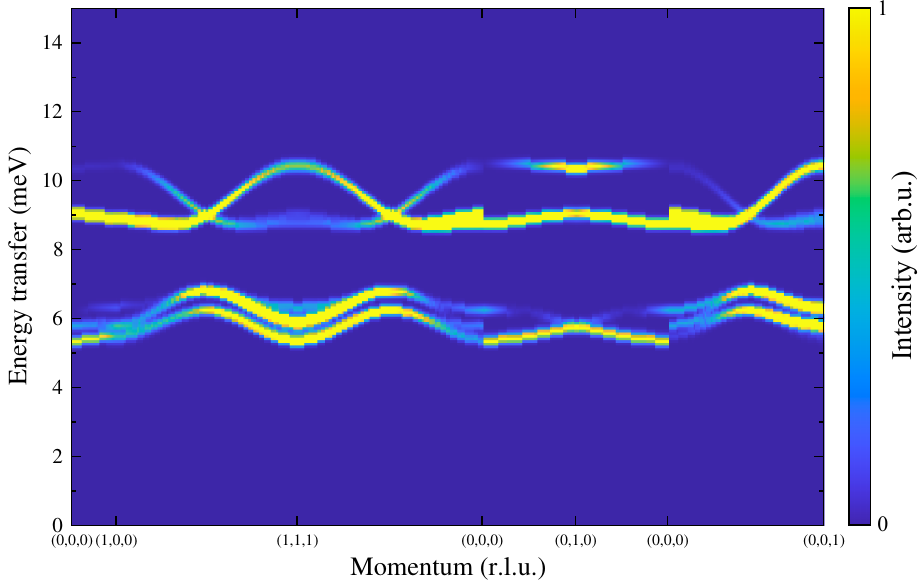}
    \caption{Dynamical structure factor for the parameters from the main text along several high-symmetry momentum directions.}
    \label{fig_SM_lswt}
\end{figure}

While we present the powder-averaged dynamical structure factor in the main text, here we also show the result for various one-dimensional cuts in the Brillouin zone, see Fig.~\ref{fig_SM_lswt}.

\end{document}